\documentclass[
floatfix,
superscriptaddress,
 showkeys,
 aps,
reprint
]{revtex4-2}

\usepackage{amsmath, amssymb}
\usepackage{gensymb}    
\usepackage{graphicx}
\usepackage{enumerate}
\usepackage{hyperref}
\usepackage{multirow}   
\usepackage{wrapfig}
\usepackage{dsfont}     
\usepackage{slashed}   
\usepackage{url}
\usepackage{siunitx}
    \let\svqty\qty
\usepackage{physics}
    \let\qty\svqty
\usepackage{xcolor}
\usepackage{chemformula}
\usepackage{mathtools}
\usepackage{bm}
\usepackage{float}
\usepackage{fancyhdr}

\newcommand{\mueV}{\mu\text{eV}}

\begin{document}

\title{First results of the SUPAX Experiment: Probing Dark Photons} 

\author{Tim Schneemann}
\affiliation{Johannes Gutenberg-Universit{\"a}t Mainz, 55128 Mainz, Germany}
\author{Kristof Schmieden}
\affiliation{Johannes Gutenberg-Universit{\"a}t Mainz, 55128 Mainz, Germany}
\author{Matthias Schott}
\affiliation{Johannes Gutenberg-Universit{\"a}t Mainz, 55128 Mainz, Germany}

\date{\today}

\begin{abstract}

We show the first results of a new cavity based haloscope searching for dark photons with masses around $34~\mueV$.
Dark photons are hypothetical vector particles and a compelling dark matter candidate. Having the same quantum numbers as photons a kinematic mixing between both is expected, leading to conversions from dark photons to standard model photons, where the photon frequency depends on the dark photon mass. 
For wavelengths in the microwave regime resonators are typically used to enhance the  signal. 
A new experiment is setup at the University of Mainz. In this paper we present the initial results from the new setup searching for dark photons utilising a 8.3 GHz copper cavity at LHe temperatures. Limits on the kinetic mixing parameter $\chi < (6.20 \pm 3.15^\text{(exp.)} \pm 9.65^\text{(SG)}) \cdot 10^{-14}$ at 95\% CL are set at a single frequency as proof of concept.
Finally the next steps of the experiment and expected sensitivity are detailed.

\end{abstract}


\maketitle

\section{Introduction}
\label{sec:intro}

\textit{Dark Photons} (DP) are hypothetical massive vector bosons with no direct coupling to Standard Model (SM) particles. 
They are compelling candidates for dark matter \cite{PhysRevD.104.095029,Fabbrichesi_2021} and as such attract increasing attention during the last decade. 
Similar to axions and axion-like particles, additional hidden sector U(1) gauge bosons are also a generic feature arising in string compactifications \cite{Jaeckel_2010} and are hence well motivated. They are represented
by a new U(1) symmetry where the Lagrangian including the dark photon with mass $m_{A'}$ is
\begin{multline}
    \mathcal{L} = - \frac{1}{4}(F^{\mu\nu}_1 F_{1 \mu\nu} + F^{\mu\nu}_2 F_{2 \mu\nu} - 2 m_{A'}^2 {A'}^2  \\ 
    - \underbrace{2 \chi F^{\mu\nu}_1 F_{2 \mu\nu} }_\text{interaction term} )
\end{multline}
where $F^{\mu\nu}_1$ is the electromagnetic (EM) field tensor and $F^{\mu\nu}_2$ is the dark photon field tensor.
As DPs have the same quantum numbers as standard model photons kinetic mixing between both is expected \cite{HOLDOM1986196}, leading to an effective coupling between DPs and SM particles. The mixing strength is giving by the kinetic mixing parameter $\chi$.

Assuming DPs are the main component of the dark matter halo in our galaxy a cavity based haloscope \cite{JAECKEL2008509} is a sensitive probe to detect dark photons. The haloscope consists of a microwave resonator at cryogenic temperatures where a converting DP will excite a radio frequency (RF) field, which is resonantly enhanced if it's resonance frequency is close to the one of the cavity. The RF power is read out by low noise amplifiers. In contrast to axion haloscopes, no magnetic field is needed to detect DPs. 

This paper presents the first results from the initial setup of the new \textit{SUPAX} experiment at Mainz. It utilizes a dedicated copper cavity at LHe temperature and a high electron mobility transistor (HEMT) based low noise amplifier readout.  

This paper is structured as follows. After a brief explanation of the expected signal in section \ref{sec:theory} the experimental setup is detailed in section \ref{sec:setup} followed by the description of the cavity in section \ref{sec:cavity}. Data analysis and results are discussed in section \ref{sec:analysis} followed by an outlook to the next stages of the experiment \ref{sec:outlook}.

\section{Theoretical Background}
\label{sec:theory}

As explained in Ref. \cite{JAECKEL2008509} dark photons can be detected through their mixing with the SM photon. If dark photons oscillate into SM photons inside a microwave cavity with a large quality factor, then a feeble EM signal accumulates inside the cavity, which can subsequently be read out.
The RF power the dark photon field excites in a RF resonator is given by \cite{signal_power_1, signal_power_2, signal_power_3, signal_power_4}
\begin{align}
    P_S^{A'} = P_0 \frac{\beta}{\beta + 1} L(f, f_0, Q_L)
    \label{eq:PS_darkphoton} \\
    P_0 = \eta \chi^2 m_{A'} \rho_{A'} V_\text{eff} Q \\
    \text{with} \hspace{0.4cm} Q =
        \begin{cases}
            Q_L &\text{if $Q_L < Q_{\text{DM}}$}\\
            Q_{\text{DM}} &\text{if $Q_L > Q_{\text{DM}}$}
        \end{cases}
    \label{eq:P0_darkphoton}
\end{align}
Here $\beta$ is the coupling coefficient of the cavity, $\eta$ an attenuation factor of the experimental setup and $Q_L$ is the loaded quality factor of the cavity.  
The mass and local density of the DPs are given by $m_{A'}$ and $\rho_{A'}$.  
$Q_\text{DM} \approx 10^6$ is the dark matter “quality factor” related to the dark matter coherence time.  
The effective volume of the cavity $V_\text{eff}$ is given by the overlap between the dark photon field $\vb{A'}(\Vec{x})$ and the induced electric field $\vb{E}(\Vec{x})$
\begin{align}
    V_\text{eff} = \frac{(\int dV \vb{E}(\Vec{x}) \cdot \vb{A'}(\Vec{x}))^2}{\int dV \epsilon_r \abs{\vb{E}(\Vec{x})}^2 \cdot \abs{\vb{A'}(\Vec{x})}^2}.
    \label{eq:V_eff}
\end{align}
\newpage
The Lorentzian term $L$ given by 

\begin{align}
    L(f, f_0, Q_L) = \left(1 + \left(Q_L \frac{f-f_0}{f_0}\right)^2 \right)^{-1}
\end{align}
limits the sensitivity of the experiment to mass ranges in close proximity to the resonant frequency of the cavity $f_0$, 
where the conversion from mass to frequency is given by ${f = {m_{A'} c^2}/{h}}$.

\section{Experimental Setup}
\label{sec:setup}

\begin{figure}[ht]
    \centering
    \includegraphics[width=0.48\textwidth]{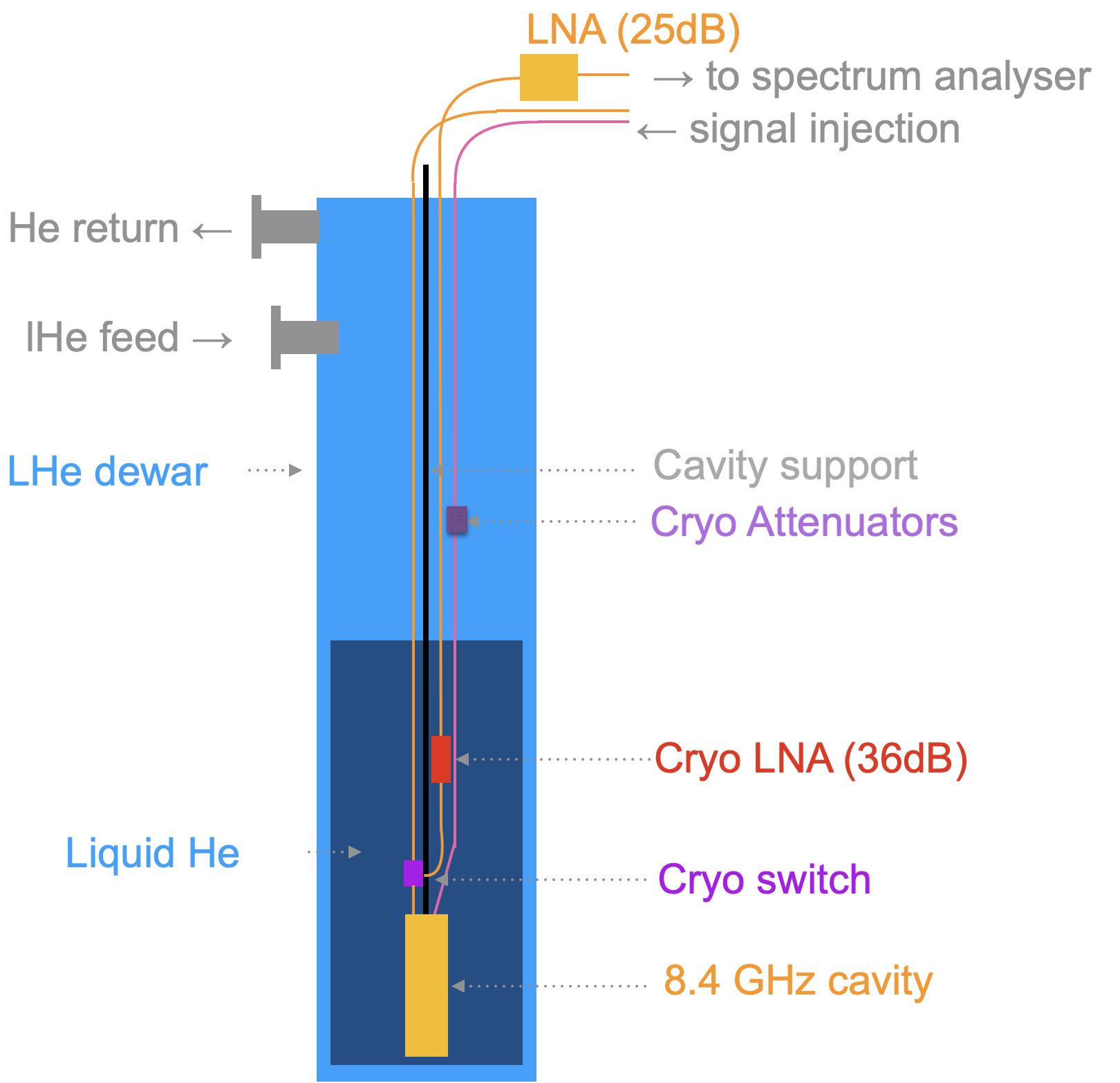}
    \caption{Scheme of setup inside the LHe dewar.}
    \label{fig:setup_scheme}
\end{figure}

\noindent 
A schematic drawing of the setup is shown in Figure \ref{fig:setup_scheme}.
The cavity has an outer size of $\approx160 \times 40 \times \SI{26.8}{mm^3}$ (for details see section \ref{sec:cavity}) and is hung from the top of a liquid helium (LHe) dewar and submerged in the LHe bath. 
The cavity has one critically and one weakly coupled port, which are connected via SMA cables. The weakly coupled antenna ($\beta \ll 1$) is used to inject signals from a Vector Network Analyser (VNA) via a \SI{20}{dB} attenuator. 
The critically coupled antenna ($\beta = 1$) is connected to a switch where one output leads straight out of the cryostat and into the VNA while the other feeds into a cryogenic low noise amplifier (LNA)\cite{LNA_datasheet} with \SI{36}{dB} gain. The amplified signal is fed into a real-time spectrum analyser (RSA) with an internal SI{25}{dB} amplifier, outside the cryostat. 

The amplifier bypass is used to measure the S-parameters of the 2 port cavity and the antenna couplings. 
During the cool down of the cavity to $T = \SI{4.2}{K}$, the resonance frequency shifts from $\SI{8.4}{GHz}$ to $\SI{8.3}{GHz}$  as the dielectric properties of the LHe overcompensate the thermal contraction of the cavity. 
The evolution of the resonance frequency is monitored until the system becomes stable after cool down. 

The RSA is connected via a USB3 link to a dedicated readout PC and controlled via custom made software. 
A 10MHz window around the center frequency is read out by the RSA and sampled with 28 MS/s. 
The real-time IQ data is streamed from the device and converted via a FFT into the frequency domain with a readout band width of 1kHz. The resulting spectra from two second intervals are averaged in real-time in stored for offline analysis.
During data taking an internal phase alignment of the RSA is triggered every 5 minutes to keep the readout stable. 
\section{RF Cavity}
\label{sec:cavity}

\subsection{Cavity Design}
\label{subsec:design}

\begin{figure}[ht]
    \centering
    \includegraphics[width=0.2\textwidth]{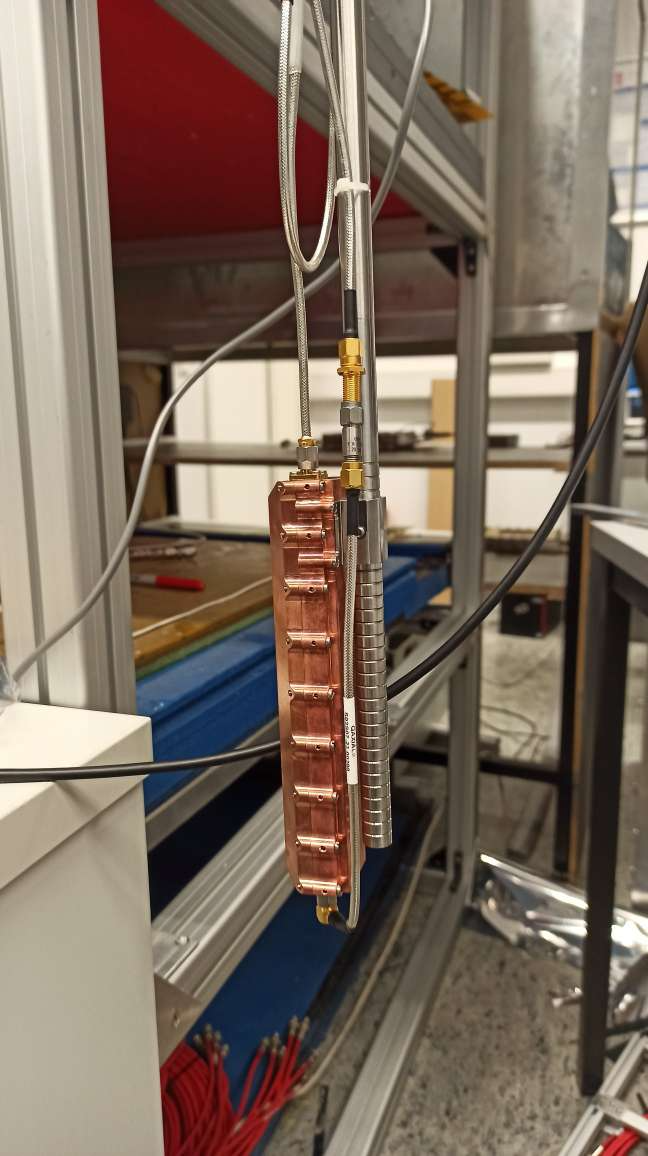}
    \caption{Final cavity used in the LHe dewar.}
    \label{fig:cavity}
\end{figure}

The cavities dimensions are determined by the combined effort to maximise the effective volume of the cavity while making sure that 
\begin{itemize}

    \item the final assembly fits into the designated cryostat in the final setup with a diameter of \SI{52}{mm},
    \item the mode next to the investigated $\text{TM}_{010}$ mode is $>\SI{100}{MHz}$ away.
\end{itemize}
These constrains yield the inner dimensions of $150 \times 22.8 \times 30 \si{mm^3}$ where the corners are rounded ($r = \SI{9}{mm}$) to reduce boundary effects. These dimensions yield a resonance frequency of
\begin{align}
    f = \SI{8.47}{GHz} \hspace{2mm} \Rightarrow \hspace{2mm} m_{A'} = 2 \pi \hbar \cdot f \approx \SI{35}{\mu eV}.
\end{align}
at room temperature. 
The cavity, shown in Figure \ref{fig:cavity} was milled into a block of solid copper with outer dimensions of $160 \times 26.8 \times \SI{40}{mm^3}$. The cavity including its holding structure weighs \SI{470}{g}, thus is relatively light.

\subsection{Cavity characterisation}
\label{subseb:characterisation}

The cavity has been characterised using a Vector Network Analyser and compared to the expected values of the S-parameters obtained by finite element simulation using the CST Studio Suite 
\cite{CST_Studio}. 
The antenna lengths have been optimised in the simulation to, one under-coupled for the RF input  and one critically coupled used for the readout. 
A comparison of the parameters obtained at room temperature and in liquid helium is shown in Table \ref{tab:Q_factors}.

\begin{table}[H]
    \centering
    \begin{tabular}{c|c|c|c|c}
         & \multicolumn{2}{c|}{Room temperature} & \multicolumn{2}{c}{Liquid He} \\
        \hline  & measurement & simulation & measurement & simulation \\
        \hline $Q_L$ & $7372 \pm 320$ & $8921 \pm 469$ & $15096 \pm 1576$ & $16277 \pm 212$ \\
        \hline $Q_0$ & $16294 \pm 707$ & $17955 \pm 943$ & -- & $39660 \pm 518$ \\
    \end{tabular}
    \caption{Comparison of measurement and simulation of the loaded and unloaded quality factors at room temperature and in liquid helium.}
    \label{tab:Q_factors}
\end{table}

The unloaded quality factor $Q_0$ was determined at $T=4\,\text{K}$ in a dedicated measurement in vacuum using a cold-head cryostat due to limited availability of the LHe setup. The measured value of $Q_0 = 40726 \pm 4250$ agrees well with the value obtained from simulation $Q_0^{\text{sim}} = 40059 \pm 128$. 
Hence the following calculations use the quality factor given by the simulation of the cavity in LHe as given in Table \ref{tab:Q_factors}, which differs slightly from the above measurement due to the shifted resonance frequency going from vacuum to LHe.
The coupling $\beta$ of the antenna is determined to be $\beta = 1.627 \pm 0.279$.
The antenna is slightly overcoupled which results in a sub-optimal sensitivity to dark photon signals.
The antenna coupling will be optimised in the following measurement runs.

\section{Data Analysis and Results}
\label{sec:analysis}

\subsection{Calibration of the readout system}
\label{subsec:gain_curve}

The readout system of the experiment is calibrated to remove any artifacts of the gain curve of the system and electronic, non-Gaussian contributions to the noise spectrum. 

\begin{figure}[ht]
    \centering
    \includegraphics[width=0.5\textwidth]{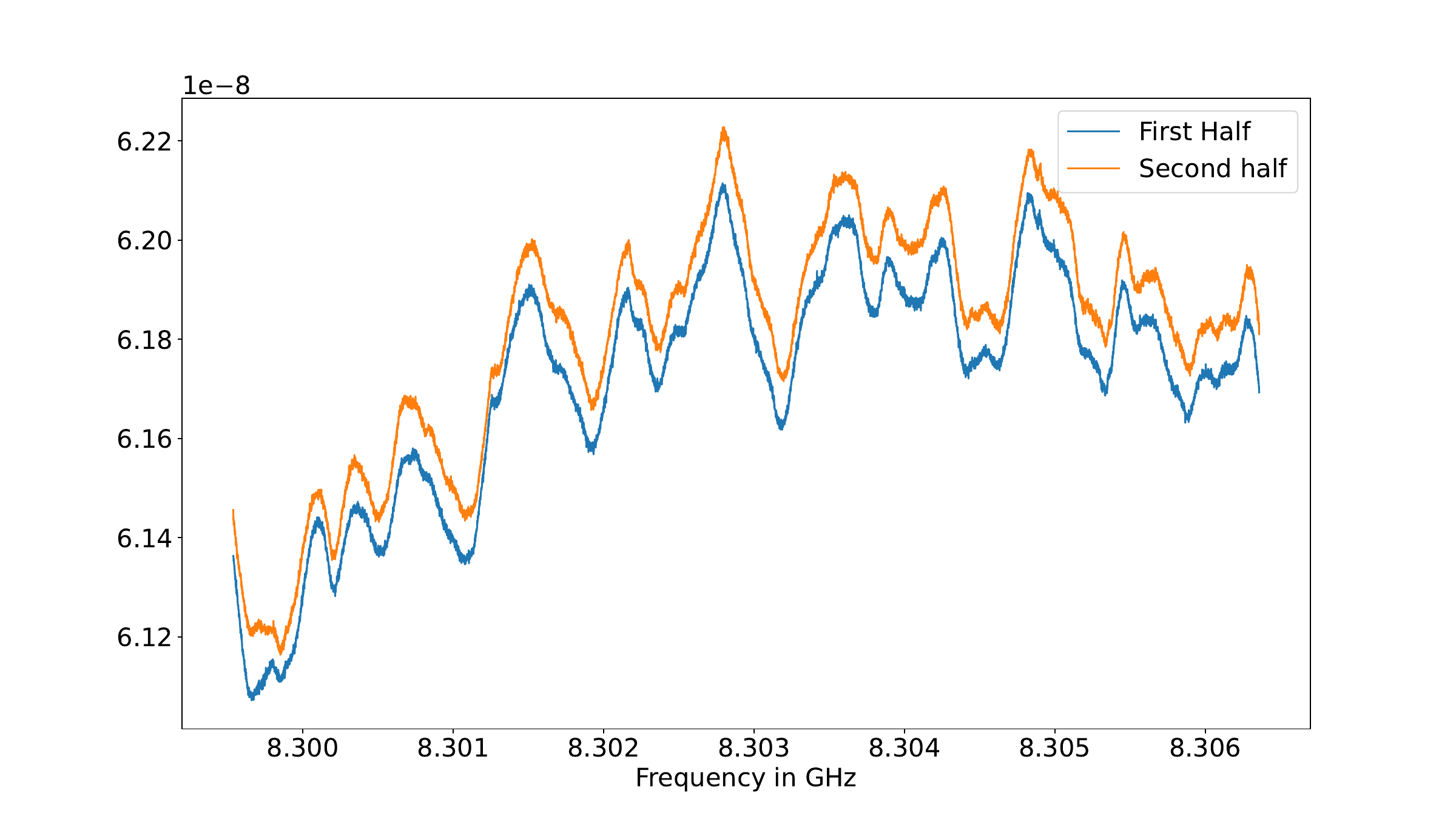}
    \caption{Hours 1-6 and 7-12 of a 12 h consecutive electronic noise (gain curve) acquisition. The difference in height is not important, since this is removable by dividing by its mean. However the slight changes in structure, as can be clearly seen e.g. in the frequency region just before \SI{8.300}{GHz}, are evidence of a the gain curve slightly varying over time, causing the need for the SG filter treatment.}
    \label{fig:gain_curve}
\end{figure}

For an ideal data acquisition system the gain curve is expected to be flat, the noise only statistical and therefore the standard deviation of noise would decrease over time by ${1}/{\sqrt{t}}$. 
The noise spectrum of the readout system, as shown in Figure \ref{fig:gain_curve}, is measured at room temperature by terminating the input with a $50\,\Omega$ resistor. The obtained noise spectrum, which is a good estimate for the gain curve of the system, is not flat but shows pronounced features. 
While the coarse structure remains constant over time, the fine structure changes, which makes a correction for the electronic structure non-trivial. As adaptive solution a Savitzky–Golay (SG) filter~\cite{SG_filter} is used to remove the electronic structure from the noise with window with of \SI{101}{kHz} and a polynomial order of $4$.

\begin{figure}[ht]
    \centering
    \includegraphics[width=0.5\textwidth]{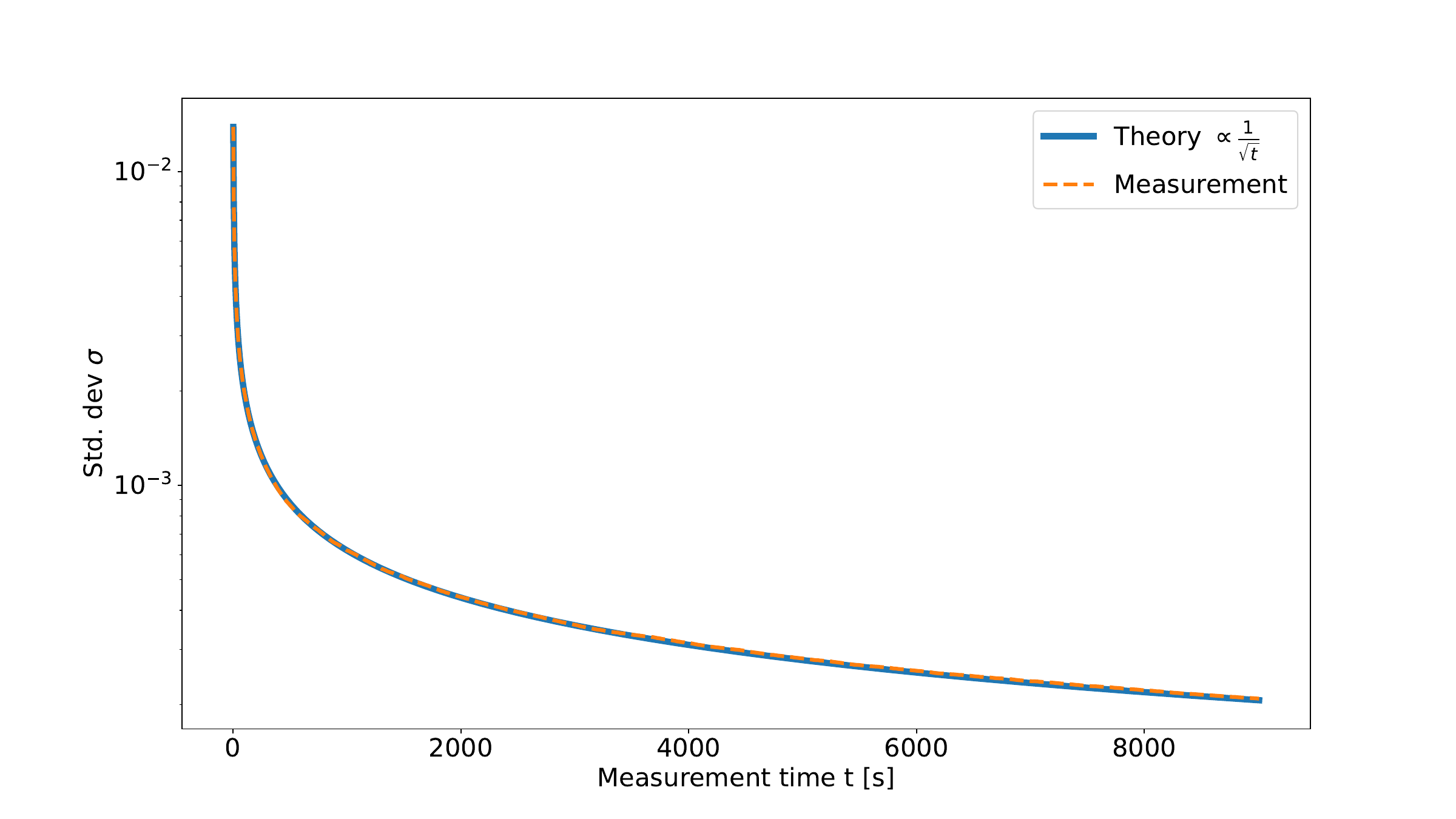}
    \caption{Behaviour of $\sigma \left(\frac{\text{gain curve}}{\text{SG filtered gain curve}}\right)$ over 6:40h against time. The magnitude of the noise around the general structure of the gain curve behaves Gaussian and therefore behaves as expected by theory.}
    \label{fig:gaussian_noise}
\end{figure}

\begin{figure}[ht]
    \centering
    \includegraphics[width=0.4\textwidth]{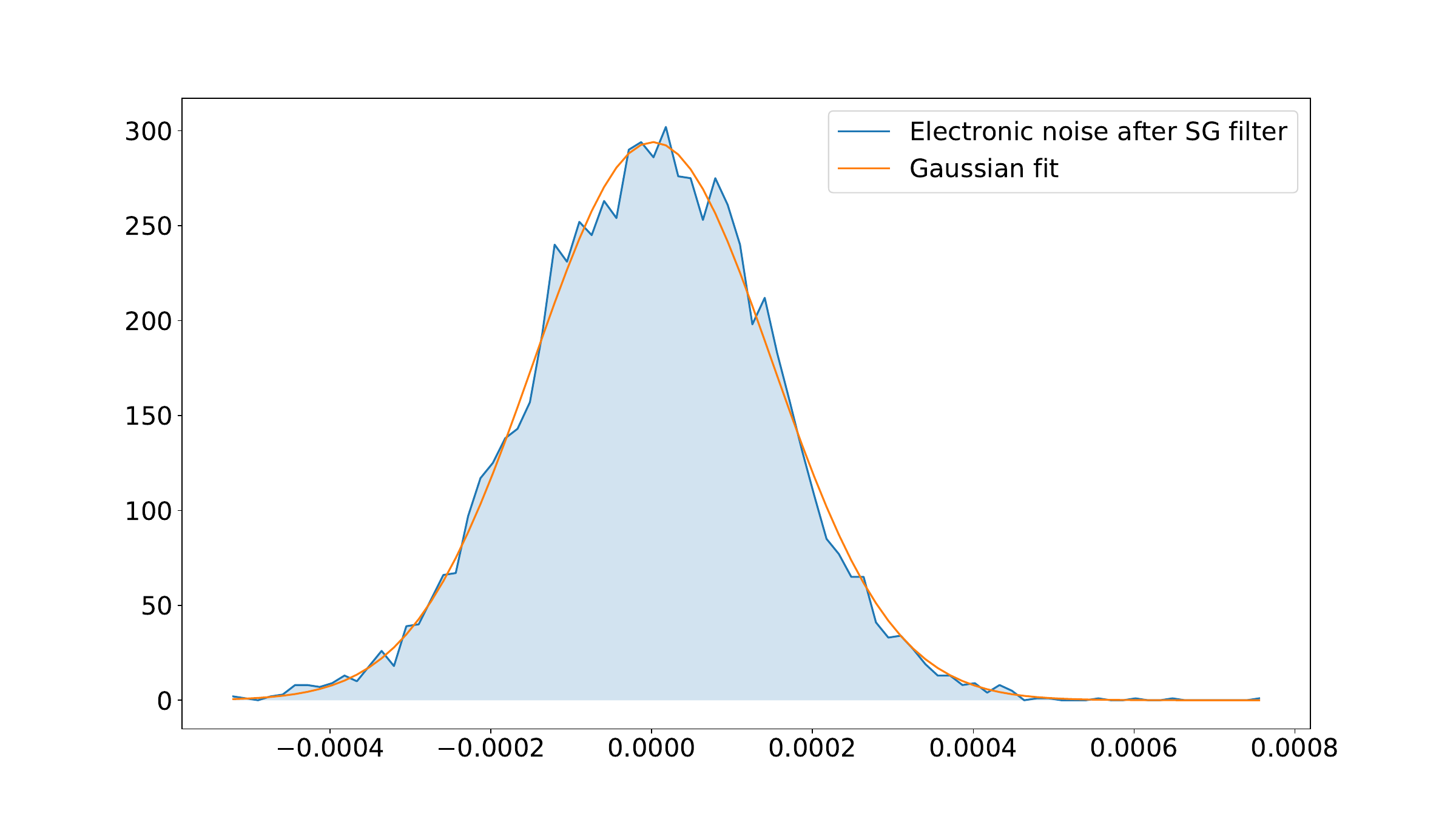}
    \caption{Histogram of the noise spectrum after integrating 6h and 40min of data. The distribution is in good agreement with a Gaussian distribution.}
    \label{fig:NoiseHistogram}
\end{figure}

After removing the electronic structure the remaining noise is expected to behave Gaussian. 
The width of the noise distribution is shown in Fig. \ref{fig:gaussian_noise} in dependence of the integration time $t$. 
It nicely follows the expected $1/\sqrt{t}$ behaviour. The resulting noise distribution, after 6h of integration, behaves perfectly Gaussian as can be seen in Figure \ref{fig:NoiseHistogram}.

\subsection{Dark photon search}
\label{subsec:DP_search}

The recorded physics data comprises 126 mins where the cavity was inside the LHe bath at a stable temperature of $\SI{4.2}{K}$.

The data analysis is performed in four steps:
\begin{itemize}
    \item Removal of the coarse structure imprinted by the electronics on the measured spectrum
    \item Removal of the variable gain curve and the cavity's resonance structure via an SG filter
    \item Normalisation of the resulting spectrum
    \item Limit setting
\end{itemize}

\begin{figure}[ht]
    \centering
    \includegraphics[width=0.5\textwidth]{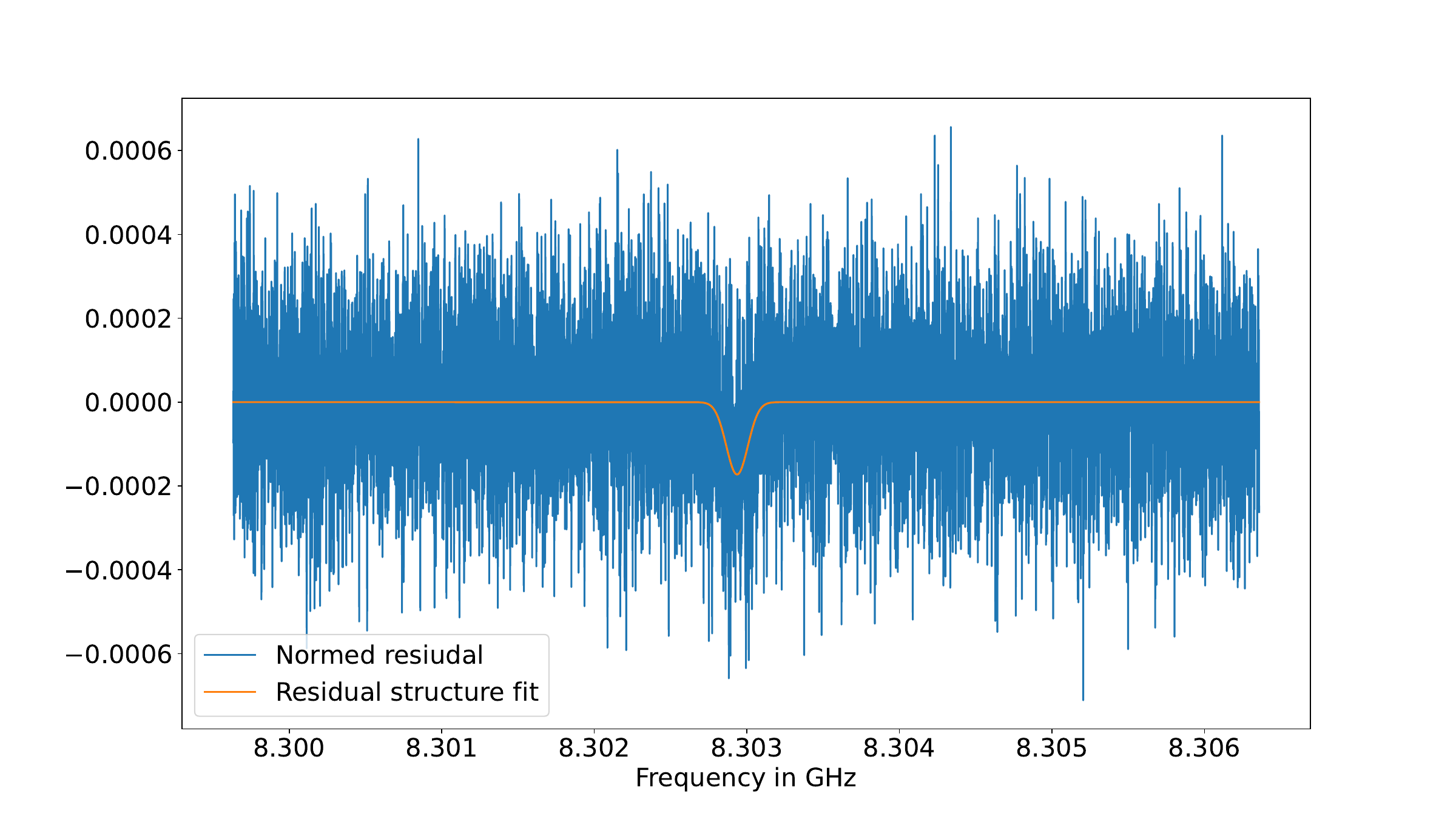}
    \caption{Residual and structure correction fit of data over SG filter(data).}
    \label{fig:residual}
\end{figure}

The coarse and stable structure of the gain curve is removed from the measured spectra by dividing by the previously measured gain curve of the readout system.
The resulting spectra are averaged in one minute intervals, i.e. $30 \times \SI{2}{s}$ spectra. Those are passed through a SG-filter with window size \SI{101}{kHz} and a polynomial order of $4$ and divided by the result. This removes the variable electronic structure as well as the cavity's resonant structure. 
To minimize the attenuation of a potential dark photon signal, which would be a narrow peak in the spectrum, the filter parameters are optimised so that the window is as large as possible using the minimal polynomial order that allows to completely remove all structure imprinted by the readout system.
The resulting spectra are averaged and shifted by $-1$, which 
is shown in Figure \ref{fig:residual}.

\begin{figure}[ht]
    \centering
    \includegraphics[width=0.5\textwidth]{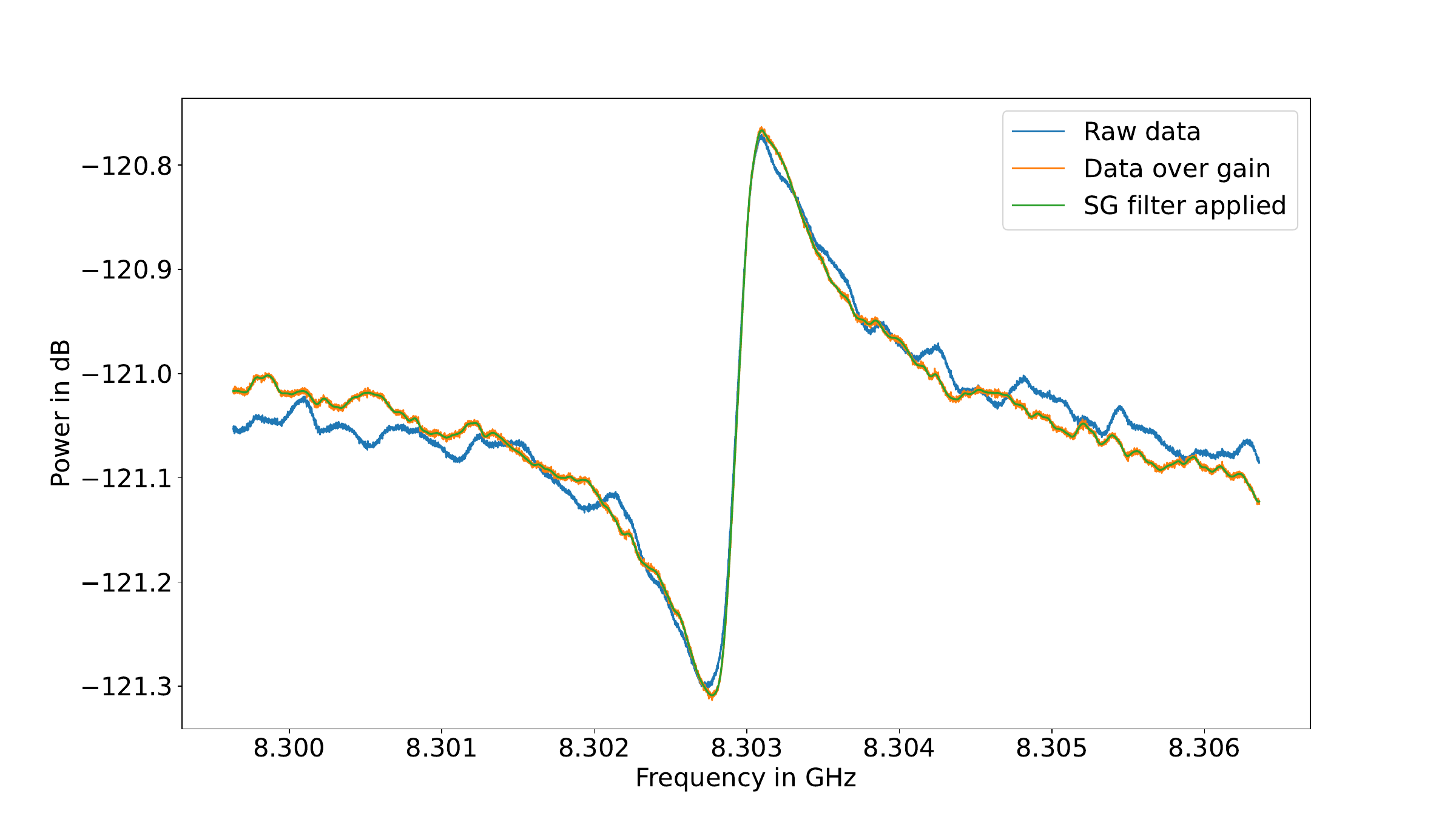}
    \caption{Raw data measured in LHe, integrated over \SI{126}{min} (blue), divided by gain curve (orange) and after the SG filter correction (green). The unusual shape of the peak is caused by reflections at the pre-amplifier and will be removed by a circulator in the final setup.}
    \label{fig:LHe_integrated}
\end{figure}

The such corrected spectrum still shows a systematic structure at the centre frequency due to the steep slope in the raw spectrum, see Figure \ref{fig:LHe_integrated}, which is not correctly captured by the SG filter. 
The structure is well described by a Gauss function which is fitted to the spectrum as shown in \ref{fig:residual} and removed by dividing the spectrum with the fit result.

Since the quality factor of a DP peak is $Q_{DM} = 10^6$ the FWHM and standard deviation of a Gaussian shaped DP peak at the cavity's resonance frequency would be
\begin{align}
    \mathrm{FWHM} = \frac{f_0}{Q_{DM}} = \SI{8.303}{kHz} \\
    \sigma = \frac{\mathrm{FWHM}}{2\sqrt{2\ln 2}} = \SI{3.53}{kHz}.
\end{align}
Hence the RF power of a DP signal would be spread over several bins in the spectrum. To increase the sensitivity a moving average with a window width of \SI{21}{kHz} is applied, corresponding to a $\pm 3\,\sigma$ interval including $\SI{99.87}{\%}$ of the the energy of a Gaussian shaped signal.

\begin{figure}[ht]
    \centering
    \includegraphics[width=0.5\textwidth]{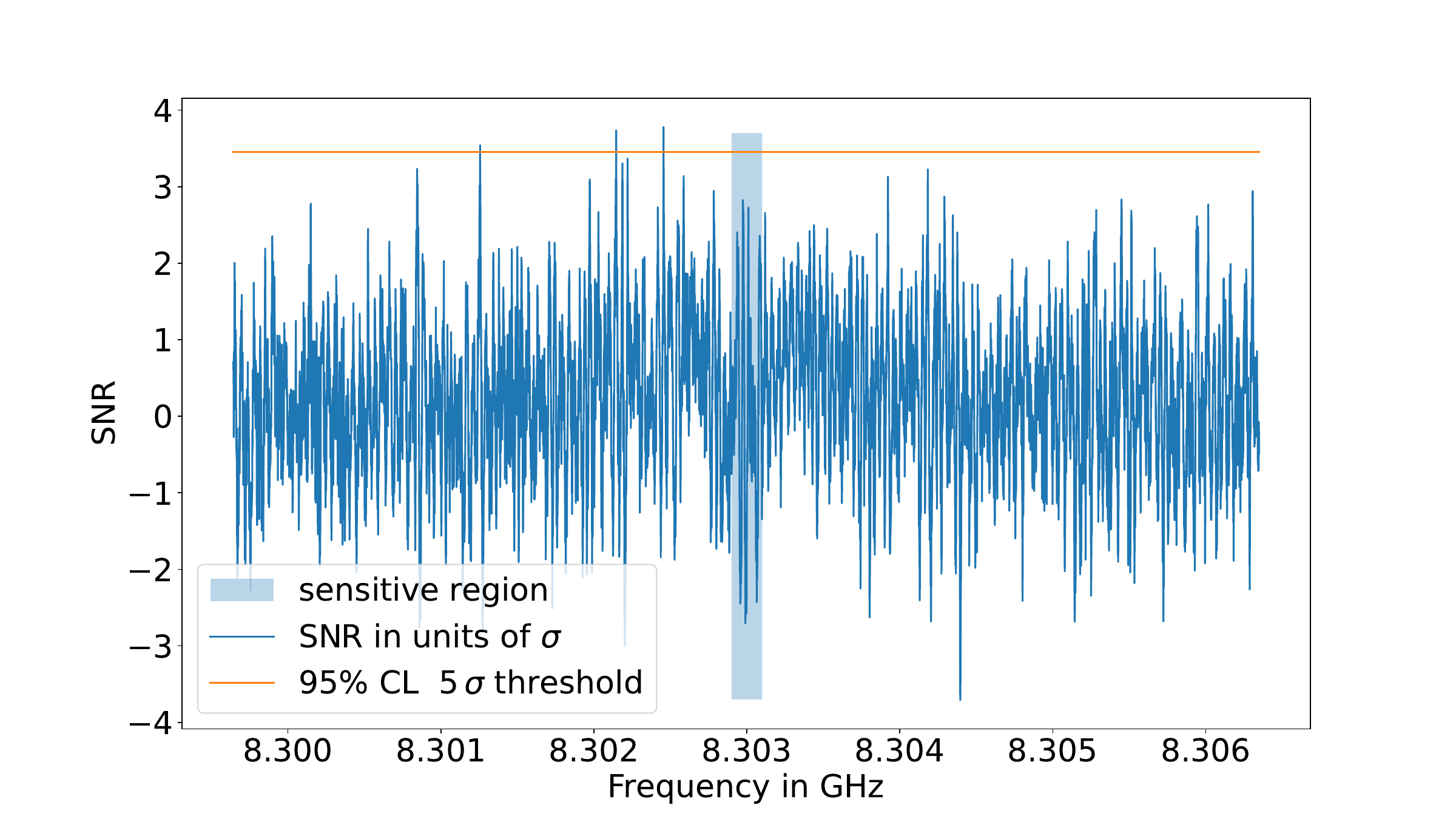}
    \caption{SNR of the \SI{126}{min} data acquisition in units of the standard deviation $\sigma$. The considered signal region is indicated as shaded area.}
    \label{fig:SNR_meas}
\end{figure}

The resulting averaged spectrum is normalized by its standard deviation to yield a measure of the signal to noise ratio (SNR) in each frequency bin in units of the standard deviation $\sigma$, as shown in Figure \ref{fig:SNR_meas}.

In summary, the SNR is calculated as
\begin{align}
    \mathrm{SNR} &= \frac{\delta}{\sigma} \\
    \delta &= \frac{\delta^\prime}{\text{gauss fit result}} \\
    \delta^\prime &= \frac{\text{data}_\text{cor}}{\text{SG filter}(\text{data}_\text{cor})} \\
    \text{data}_\text{cor} &= \frac{\text{measured data}}{\text{measured gain curve}} \\
    \sigma &= \sqrt{\frac{\Sigma (\delta_i - \bar{\delta})^2}{n-1}} \text{   for } i = 0,...,n
\end{align}
where $n$ denotes the number of data points in the raw data spectrum and $\delta$ is the fully corrected residual spectrum.

As shown in Ref. \cite{HAYSTAC_analysis} the absence of any signal in the SNR above $3.455 \, \sigma$ corresponds to a $95\%$ CL exclusion of the mixing parameter $\chi$, where the associated expected SNR is $5 \, \sigma$. 
For the presented measurement no signal above this limit is observed within the FWHM of the cavity peak, which is considered the sensitive region. Hence a limit is set on the kinematic mixing parameter $\chi$:
\begin{equation}
    \chi^\text{limit} = \chi_\text{guess} \sqrt{\frac{\text{SNR}_\text{meas}}{\text{SNR}_\text{exp}}},
\end{equation}
where $\text{SNR}_\text{exp} \propto \chi^2_\text{guess}$ and $\text{SNR}_\text{meas} \propto \chi^2_{\text{DP}}$.
The measured $\text{SNR}_\text{meas}$ is rescaled by the expected $\text{SNR}_\text{exp} = \sqrt{\Delta\nu\tau}\cdot P_S / P_n$.  The signal power $P_S$, given in eq. (\ref{eq:PS_darkphoton}), uses a freely chosen value of $\chi_\text{guess} = 10^{-14}$. The expected noise power $P_n$ is given by 
\begin{align}
    P_n &= k_B T_\text{sys} \Delta\nu,
\end{align}
where  $\tau = \SI{126}{min}$ is the measurement time and  $\Delta\nu = \SI{1}{kHz}$ the frequency bin width. The parameters used in the calculation are summarized in Table \ref{tab:ParameterValues}.

\begin{table}[ht]
    \centering
    \begin{tabular}{c|c}
    Parameter & Value \\
    \hline
         $\rho_{DM}$ & 0.45 ${\si{GeV}}/{\si{cm}^3}$ \\
         $V_\text{eff}$ & $18.76 \pm 0.17 \, \si{cm^3}$ \\
         $\eta$ & $0.89 \pm 0.05$ \\
         $\beta$ & $1.627 \pm 0.279$ \\
         $Q_L$ &  $15096 \pm 1576$  \\
         $\nu$ & $\SI{1}{kHz}$ \\
         $\tau$ & $\SI{126}{min}$\\
    \end{tabular}
    \caption{Parameters used in the caluclation of the expected signal and noise.}
    \label{tab:ParameterValues}
\end{table}

For the local dark matter density $\rho_{DM}$ the commonly used value from \cite{PhysRevD.104.095029} is adopted. $m_{A'}$ is set by the resonance frequency of the cavity in the liquid helium environment. 
The effective volume is calculated using eq.~(\ref{eq:V_eff}) in the CST Studio simulation with the uncertainty originating in the precision of the cavity milling of $\pm \SI{0.01}{mm}$ in every dimension. 
The signal attenuation factor $\eta$ is composed of a $\SI{0.13}{dB}$ signal loss due to \SI{10}{cm} of cable between the antenna and the cryo LNA and an additional signal loss of \SI{3.6}{dB} after the LNA.
$\eta$ can be further optimised by shortening the cable distance between the cavity antenna and the cryo LNA. An uncertainty of $5 \%$ is assumed on the attenuation factor to account for losses in SMA connectors used after the LNA.

\begin{figure}
    \centering
    \includegraphics[width=0.48\textwidth]{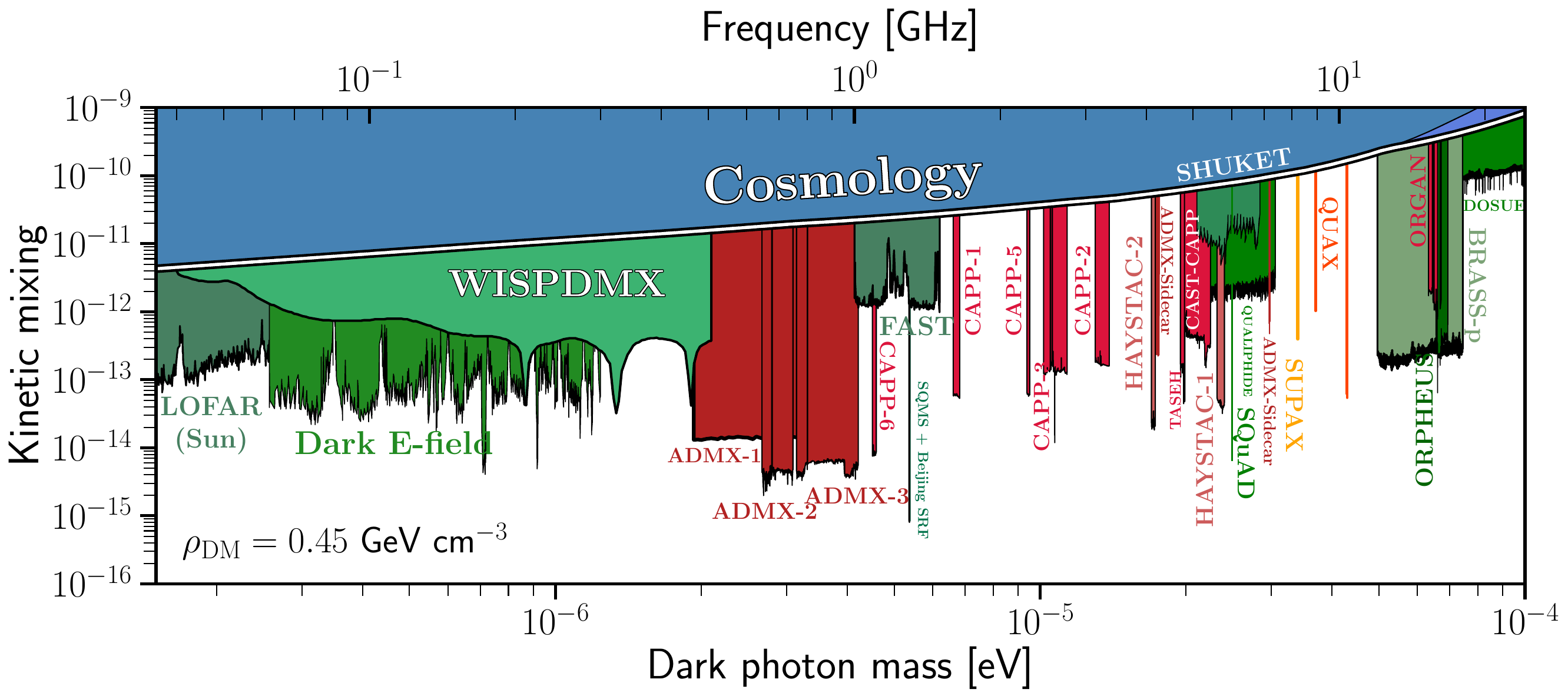}
    \caption{Limit of the presented experiment in comparison to other experiments, adapted from Ref. \cite{AxionLimits}. }
    \label{fig:Limits}
\end{figure}

At the resonance frequency of the cavity $f_0 = \SI{8303.06}{MHz} \equiv 34.34 \, \mu$eV a limit of \begin{align}
    \chi_\text{limit} < (1.24 \pm 0.63^\text{(exp.)} \pm 1.93^\text{(SG)}) \cdot 10^{-14}
\end{align} is calculated, assuming a randomly polarised dark photon field. 
The analysis is validated by injecting a synthetic DP signal with $Q_{DM} = 10^6$ and $\chi = n\cdot\chi_\text{limit}$ into the raw data stream. 
The resulting signal is reliably seen well above the threshold of $\text{SNR} > 3.455\sigma$ only when increasing $n$ to at least 5, indicating unaccounted losses in the analysis procedure attenuating a signal by $80\%$. Clearly the analysis procedure will be revisited in the future. This leads to our final claimed limit of 
\begin{align}
    \chi_\text{limit} < (6.20 \pm 3.15^\text{(exp.)} \pm 9.65^\text{(SG)}) \cdot 10^{-14}
\end{align}
In the case of a DP field with fixed polarisation, the sensitivity is reduced by a factor of $\sim 6$ for a two hour measurement, as calculated in Ref. \cite{DP_handbook}. This translates into a limit in of $\chi_\text{pol.} < (38 \pm 19^\text{(exp.)} \pm 58^\text{(SG)}) \cdot 10^{-14}$, 
which is already in the same ball park as some measurements in this frequency range \cite{PhysRevD.104.095029}. Our result is summarized together with other experiments in Figure \ref{fig:Limits}. 

The experimental  uncertainty (exp.) includes all uncertainties from the characterisation of the cavity and from the readout. 
The uncertainty of the SG filter has been determined by changing the order of the polynomial fit and the width of the moving window. 
The loss of sensitivity is most probably to be attributed to a significant signal attenuation by the SG filter in the analysis. 
To test this it is planned to inject a physical signal into the cavity during a measurement run and to check how well the analysis performs on this data.

\section{Conclusion and Outlook}
\label{sec:outlook}

Approximately one year after the idea of the SUPAX experiment we can report on the successful design and construction of an RF-cavity, the entire data acquisition hard- and software at room and cryogenic temperature. The copper cavity has a comparable unloaded quality factor as similar experiments like RADES and CAPP. 
Furthermore, although the experiment is originally aimed at detecting axions, an approximately 2h long data taking run in the absence of a magnetic field has been conducted and analysed with the goal to search for dark photons, another dark matter candidate. 
In the absence of a signal
an upper limit is obtained for the kinetic mixing parameter of $\chi < (6.20 \pm 3.15^\text{(exp.)} \pm 9.65^\text{(SG)}) \cdot 10^{-14}$ for a dark photon mass of $m_{A'} = 34.34 \, \mu$eV in the randomised polarisation scenario.

There are several future developments for the experiments planned, which will increase the sensitivity of SUPAX to the same level as the most sensitive experiments in this frequency range: 
an identical cavity is currently being coated with a superconductor which is yet to be tested in the context of RF cavities. It is to be studied how the superconductor performs and if it can improvement the unloaded quality factor in the magnetic field. 
We expect $Q_0$ to be around few times $10^5$, increasing the limit by one order of magnitude. 
In addition we expect upgrades on the setup. Introducing a circulator will avoid any RF reflections seen in the previous data taking. The temperature will be lowered to 1.5K and the now installed cryostat will allow for longer measurement times. 

The 14T magnet at the Helmholz Institute Mainz will become operational within the second half of 2023, allowing for a first search for axion-like particles with SUPAX.

\section{Acknowledgement}
\label{sec:acknoledgement}

We thank Jessica Golm for many helpful discussions and for providing the initial cavity design. We also thank the members of the RADES collaboration for their valuable insights into the RF readout and discussions. 
This work would have not been possible without the ERC-Grant “LightAtTheLHC” as well as the continuous support from the PRISMA+ Cluster of Excellence at the University of Mainz.

\bibliographystyle{ieeetr}
\bibliography{Supax_DP}

\end{document}